\documentclass{jfm}

\usepackage{graphicx}
\usepackage{natbib}
\usepackage{color}

\ifCUPmtlplainloaded \else
  \checkfont{eurm10}
  \iffontfound
    \IfFileExists{upmath.sty}
      {\typeout{^^JFound AMS Euler Roman fonts on the system,
                   using the 'upmath' package.^^J}%
       \usepackage{upmath}}
      {\typeout{^^JFound AMS Euler Roman fonts on the system, but you
                   dont seem to have the}%
       \typeout{'upmath' package installed. JFM.cls can take advantage
                 of these fonts,^^Jif you use 'upmath' package.^^J}%
      }
  \else
  \fi
\fi

\ifCUPmtlplainloaded \else
  \checkfont{msam10}
  \iffontfound
    \IfFileExists{amssymb.sty}
      {\typeout{^^JFound AMS Symbol fonts on the system, using the
                'amssymb' package.^^J}%
       \usepackage{amssymb}%

      }{}
  \fi
\fi

\ifCUPmtlplainloaded \else
  \IfFileExists{amsbsy.sty}
    {\typeout{^^JFound the 'amsbsy' package on the system, using it.^^J}%
     \usepackage{amsbsy}}
    {\providecommand\boldsymbol[1]{\mbox{\boldmath $##1$}}}
\fi

\newsavebox{\astrutbox}
\sbox{\astrutbox}{\rule[-5pt]{0pt}{20pt}}

\newcommand\be{\begin{equation}}
\newcommand\ee{\end{equation}}
\newcommand{\bea}{\begin{eqnarray}}
\newcommand{\eea}{\end{eqnarray}}
\newcommand\bes{\begin{equation*}}
\newcommand\ees{\end{equation*}}
\newcommand{\pdv}[2]{\frac{\partial#1}{\partial#2}}
\newcommand{\dv}[2]{\frac{d#1}{d#2}}

\newcommand{\rhob}{{\rho}}
\newcommand{\pb}{{p}}

\newcommand{\vrp}{v_r^\prime}

\newcommand{\rhop}{\rho^\prime}

\newcommand{\blds}[1]{\boldsymbol{#1}}

\title[Stability of homologous implosions]{Buoyancy instability of homologous implosions}

\author[B. M. Johnson]%
{B. M. Johnson$^1$%
  \thanks{Email address for correspondence: johnson359@llnl.gov}}

\affiliation{$^1$Lawrence Livermore National Laboratory, Livermore,
CA 94550, USA}

\pubyear{2010}
\volume{650}
\pagerange{119--126}
\date{2 April 2015; revised 19 May 2015; accepted 28 May 2015}
\begin{document}

\maketitle

\begin{abstract}
I consider the hydrodynamic stability of imploding ideal gases as an idealized model for inertial confinement fusion capsules, sonoluminescent bubbles and the gravitational collapse of astrophysical gases. For oblate modes (short-wavelength incompressive modes elongated in the direction of the mean flow), a second-order ordinary differential equation is derived that can be used to assess the stability of any time-dependent flow with planar, cylindrical or spherical symmetry. Upon further restricting the analysis to homologous flows, it is shown that a monatomic gas is governed by the Schwarzschild criterion for buoyant stability. Under buoyantly unstable conditions, both entropy and vorticity fluctuations experience power-law growth in time, with a growth rate that depends upon mean flow gradients and, in the absence of dissipative effects, is independent of mode number. If the flow accelerates throughout the implosion, oblate modes amplify by a factor $(2C)^{\left|{\color{black}N_0}\right| t_i}$, where $C$ is the convergence ratio of the implosion, {\color{black}$N_0$ is the initial buoyancy frequency} and $t_i$ is the implosion time scale. If, instead, the implosion consists of a coasting phase followed by stagnation, oblate modes amplify by a factor $\exp\left(\pi \left|{\color{black}N_0}\right| t_s\right)$, where {\color{black}$N_0$ is the buoyancy frequency at stagnation} and $t_s$ is the stagnation time scale. Even under stable conditions, vorticity fluctuations grow due to the conservation of angular momentum as the gas is compressed. For non-monatomic gases, this additional growth due to compression results in weak oscillatory growth under conditions that would otherwise be buoyantly stable; this over-stability is consistent with the conservation of wave action in the fluid frame. The above analytical results are verified by evolving the complete set of linear equations as an initial value problem, and it is demonstrated that oblate modes are the fastest-growing modes and that high mode numbers are required to reach this limit (Legendre mode $\ell \gtrsim 100$ for spherical flows). {\color{black}Finally, comparisons are made with a Lagrangian hydrodynamics code, and it is found that a numerical resolution of ${\sim}30$ zones per wavelength is required to capture these solutions accurately. This translates to an angular resolution of ${\sim} (12/\ell)^\circ$, or $\lesssim 0.1^\circ$ to resolve the fastest-growing modes.}
\end{abstract}


\section{Introduction}

Examples of imploding gases include inertial confinement fusion (ICF) capsules \citep{Atzeni04}, sonoluminescent bubbles \citep{Suslick08} and core-collapse supernovae \citep{Janka12}. The hydrodynamic stability of these flows is an important issue, as perturbations of sufficient amplitude can drain the energy driving the implosion, and the breakdown of symmetry, even for small amplitudes, can have important diagnostic/observational effects \citep[e.g.][]{Murphy14}. Stability analyses of an imploding gas are significantly complicated by the time-dependent mean flow, and homologous flow (i.e. flow in which fluid elements share a common time-dependent scaling, the Hubble flow being a prime example) is a useful idealization that allows some analytical progress to be made. Such a study was performed for core-collapse supernovae by \cite{Goldreich80}, who claim stability, but their results have recently been revisited by \cite{Cao09,Cao10}, who claim instability. \cite{Chu96} applied the analysis of \cite{Goldreich80} to a sonoluminescing bubble and also claimed stability. The extensive body of work on the stability of ICF implosions has focused almost exclusively on interfacial instability, with little attention being given to the stability of the gas (see, however, \citealt{Greenspan63,Mjolsness78,Cook00}). At the same time, hot-spot turbulence has recently received attention as a potential source of yield degradation in ICF capsules \citep{Thomas12,Gatu13,Clark13,Cerjan13}, and the origin of these vortical flows, if present, remains unclear. {\color{black}\cite{Basko98} considered the possibility of buoyancy instability in self-similar implosions but did not perform a formal stability analysis.}

The purpose of this work is to clarify some of these issues by performing a stability analysis that complements previous work. {\color{black}By solving the initial value problem \citep{Lai00}, rather than decomposing perturbations into radial modes, I am able to obtain results that are more physically transparent than those of previous authors, including a precise stability criterion.} The following assumptions are made: (i) the implosion is externally driven rather than driven by gravity; (ii) dissipative effects are ignored (this is almost certainly unrealistic for ICF capsules and bubbles, but it permits isolation of the instability driving mechanism); and (iii) perturbations are assumed to be short-wavelength, incompressive and elongated in the direction of the implosion. I shall refer to this final approximation as the oblate limit, and it is perhaps the most useful new result to come out of this work, as it allows for significant analytical progress to be made in a problem that must otherwise be treated numerically. Under these assumptions, I demonstrate that homologously imploding ideal gases are essentially governed by the Schwarzschild criterion for buoyant stability \citep{Schwarzschild92}, with a slight modification due to compression.

The basic equations and mean flow are outlined in \S\ref{sec:equations}, the stability analysis is given in \S\ref{sec:solutions}, and \S\ref{sec:discussion} summarizes and discusses applications.
 
\section{Basic equations and mean flow}\label{sec:equations}

The fundamental equations used are the continuity equation, Euler's equation, and the first law of thermodynamics for adiabatic flow:
\be\label{CONT}
\dv{\rho}{t} + \rho {\blds{\nabla}} \cdot {\blds{v}} = 0,\;\; \rho \dv{\blds{v}}{t} + {\blds{\nabla}} p = 0,\;\;\dv{s}{t} = 0.
\ee
Here $\rho$, $\blds{v}$, $p$ and $s$ are the mass density, velocity, pressure and entropy of the gas and $d/dt = \partial/\partial t + \blds{v}\cdot \blds{\nabla}$ is the Lagrangian derivative following a fluid element \citep[e.g.][]{Landau87}. I will assume throughout an ideal-gas equation of state, $p = \left(\gamma - 1\right) \rho C_v T$, where $\gamma$, $C_v$ and $T$ are the adiabatic index, specific heat and temperature of the gas. For this equation of state, $s = \ln\left(p\rho^{-\gamma}\right)$ to within a constant factor; this expression will be used in what follows to define the entropy.

Fluid elements in a homologous flow obey the relationship $r(t,a) = r(0,a) h(t) = a r_{01} h(t)$, where $r$ is the spatial coordinate in the direction of the mean flow and the Lagrangian label for a fluid element is $a \equiv r(0,a)/r_{01}$. For a spherical flow, $r(0,a)$ is the field of radial positions for all the fluid elements at $t = 0$ and $r_{01} \equiv r(0,1)$ is the outer radius of the gas at $t = 0$; this implies that the scale factor $h(t)$ is normalized to unity at $t = 0$, i.e. $h(0) \equiv 1$. The velocity field associated with homologous flow is $v_r(t,a) = a r_{01} \dot{h}(t) = r(t,a)\dot{h}/h$, where an overdot denotes a time derivative. Mass and entropy conservation under spherical adiabatic homologous flow imply $\rho(t,a) = \rho_0(a)/h^3$ and $p(t,a) = p_0(a)/h^{3\gamma}$, where a zero subscript denotes a spatial profile at $t = 0$, and Euler's equation becomes
\be\label{EULH}
h^{3\gamma-2}\ddot{h} = -\frac{1}{\rho_0 r_{01}^2 a}\pdv{p_0}{a} \equiv \pm \frac{1}{t_c^2},
\ee
where the plus (minus) sign is associated with a decelerating (accelerating) implosion and $t_c$ is a characteristic time scale.

For $\gamma = 5/3$ and $\dot{h}_0 = 0$, the time-dependent portion of expression (\ref{EULH}) yields the well-known Kidder implosions \citep{Kidder74,Atzeni04}
\be\label{KIDDH}
h(t) = \sqrt{1 \pm \left(\frac{t}{t_c}\right)^2}.
\ee
For $\dot{h}_0 \neq 0$, an additional solution to (\ref{EULH}) valid for any $\gamma$ is \citep{Goldreich80}
\be\label{HPL}
h(t) = \left(1 - \sqrt{\frac{\left[3\gamma - 1\right]^2}{6[\gamma-1]}} \frac{t}{t_c}\right)^\frac{2}{3\gamma-1}.
\ee 
This solution, while valid only for accelerating implosions, is useful for assessing the impact of the gas equation of state. Following \cite{Atzeni04}, accelerating solutions are defined as starting out from $t = 0$, so that they evolve from $h = 1$ to $h = C^{-1}$, where $C \equiv r_{\rm initial}/r_{\rm final}$ is the convergence factor, i.e, the ratio of initial to final sizes. The decelerating solution, on the other hand (the Kidder implosion with the $+$ sign), is defined as stagnating at $t = 0$, so that its implosion phase evolves from $h = C$ to $h = 1$.

\begin{figure}
\hspace{0.8in}
\includegraphics[scale=0.6]{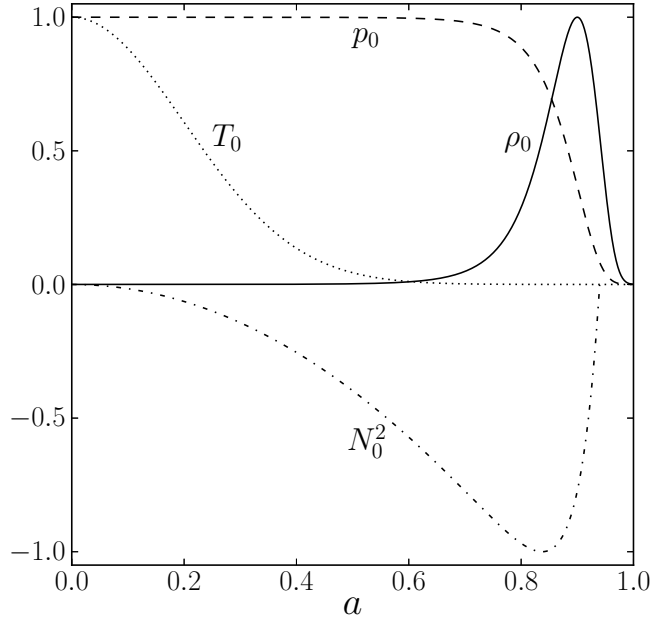}
\caption{\label{F1}Profiles of density (\emph{solid}), temperature (\emph{dotted}), pressure (\emph{dashed}) and the negative portion of $N_0^2$ (\emph{dot-dashed}) for expressions (\ref{RHOHS})--(\ref{N20HS}) with $\sigma = 0.2$ and $a_p = 0.9$. All quantities have been normalized to their peak magnitude.}
\end{figure}

Aside from the equation of state, the space-dependent portion of equation (\ref{EULH}) is the only constraint on the spatial profile of the mean gas quantities. This allows for a significant amount of flexibility in setting up various mean flow profiles. An analytical model that mimics the hot spot of an ICF implosion during deceleration (this is likely to be a good approximation for a collapsing bubble as well) is given by
\be\label{RHOHS}
\rho_0 = \rho_{0p} e^\frac{a^2 - a_p^2}{2\sigma^2}\exp\left(1-e^{\frac{a^2-a_p^2}{2\sigma^2}}\right),
\ee
\be\label{THS}
T_0 = T_{0p}e^{-\frac{a^2-a_p^2}{2\sigma^2}}, \;\; p_0 = p_{0p}\exp\left(1-e^{\frac{a^2-a_p^2}{2\sigma^2}}\right),
\ee
where the gas quantities have been normalized to their value at $a_p$ (the location of the density peak; see Fig.~\ref{F1}). Both $a_p$ and $\sigma$ can be regarded as free parameters of the model, and can be alternatively expressed in terms of ratios of physical quantities. The characteristic time scale in this case is $t_c = \sqrt{\gamma} r_{01} \sigma/c_{0p}$, where $c_{0p}$ is the sound speed at $a_p$. The square of the Brunt-V$\ddot{\rm{a}}$is$\ddot{\rm{a}}$l$\ddot{\rm{a}}$, or buoyancy, frequency,
\[
N^2 \equiv -\frac{1}{\gamma \rho}\pdv{p}{r}\pdv{s}{r},
\]
is useful for assessing buoyant stability \citep{Johnson05}. For homologous flow, $N^2 = N_0^2 h^{1-3\gamma}$, and for the hot-spot configuration described above,
\be\label{N20HS}
N_0^2 t_c^2 = -\frac{a^2}{\sigma^2}\left(1 - \frac{\gamma - 1}{\gamma} e^{\frac{a^2-a_p^2}{2\sigma^2}}\right).
\ee
Mean flow profiles for the hot-spot configuration are shown in Fig.~\ref{F1}.

\section{Stability analysis}\label{sec:solutions}

For an equilibrium configuration ($\blds{v} = 0$), it is well known that the local dispersion relation for short-wavelength incompressive modes governed by equations (\ref{CONT}) is
\be\label{DR}
\omega^2 = \frac{k_\perp^2}{k_r^2 + k_\perp^2} N^2,
\ee
where $\omega$ is the wave frequency and $k_\perp$ and $k_r$ are wavenumbers perpendicular and parallel to the mean gradients \citep[see e.g.][]{Johnson05}. It can be seen from expression (\ref{DR}) that $N^2 < 0$ denotes instability; $N^2 > 0$ is the Schwarzschild criterion for buoyant stability \citep{Schwarzschild92}. The buoyancy frequency can be rewritten as
\[
N^2 = \frac{g}{\gamma}\pdv{s}{r} = -g\left(\frac{g}{c_s^2} + \frac{1}{\rho} \pdv{\rho}{r}\right) = \frac{g}{\gamma}\left(\frac{1}{T}\pdv{T}{r} - \frac{\gamma-1}{\rho} \pdv{\rho}{r}\right),
\]
where $g \equiv dv_r/dt$ is the acceleration and $c_s$ is the mean sound speed. It can be seen from this expression that configurations with ``order-over-disorder'' ($\partial s/\partial r < 0$) are buoyantly unstable. If density gradients dominate, a ``heavy-over-light'' configuration is unstable ($\partial \rho/\partial r > 0$), whereas if temperature gradients dominate, a ``cold-over-hot'' configuration is unstable ($\partial T/\partial r < 0$). The former situation represents Rayleigh-Taylor instability, and the latter represents thermal convection; in general, buoyancy instability depends upon the mean entropy gradient. {\color{black}Because fluid elements evolve adiabatically, and the relationship between entropy and density depends upon $\gamma$, the stability criterion in general depends upon the compressibility of the fluid. Notice that in the limit $gL_\rho \ll c_s^2$, where $L_\rho = (\partial \ln \rho/\partial r)^{-1}$ is the density gradient length scale, the growth rate is independent of $\gamma$ and reduces to the short-wavelength limit of the classical Rayleigh-Taylor growth rate when the effects of density gradient stabilization are included \cite[e.g.][]{Betti98}.} For the hot-spot configuration described above, it can be seen from Fig.~\ref{F1} that the entire gas region and most of the shell is buoyantly unstable during deceleration.

For a spherically symmetric mean flow ($\blds{v} = v_r \hat{\blds{r}}$), perturbations can be decomposed into spherical harmonics $Y_{\ell m}\left(\theta,\phi\right)$, and the resulting equations are
\be\label{LR}
\dv{\rhop}{t} = -\frac{1}{r^2}\pdv{}{r}\left(r^2 v_r\right)\rhop - \frac{1}{r^2}\pdv{}{r}\left(\rhob r^2 \vrp\right) +  \frac{\ell\left(\ell + 1\right)}{r} \rhob v_\perp^\prime,
\ee
\be\label{LVR}
\dv{\vrp}{t} = - \pdv{v_r}{r} \vrp + \frac{1}{\rhob^2}\pdv{p}{r}\rhop - \frac{1}{\rhob}\pdv{p^\prime}{r},
\ee
\be\label{LVP}
\dv{v_\perp^\prime}{t} = -\frac{v_r}{r}v_\perp^\prime  - \frac{p^\prime}{\rhob r},\;\; 
\dv{s^\prime}{t} = - \pdv{s}{r}\vrp,
\ee
where a prime denotes a fluctuation, $v_\perp^\prime$ is the component of the velocity fluctuation parallel to $\blds{\nabla} Y_{\ell m}$, $s^\prime = p^\prime/p-\gamma \rhop/\rhob$ is the entropy fluctuation, and $\ell$ is a Legendre mode number \citep{Lai00}. The third velocity component decouples from the other variables and will not be considered here. Short-wavelength, low-frequency fluctuations satisfy $\blds{\nabla} \cdot {\textbf {\emph v}}^\prime \approx 0$ and $s^\prime \approx -\gamma \rhop/\rhob$ (the Boussinesq approximation for buoyancy-driven flows; \citealt{Johnson05}). This reduces the perturbed continuity equation to the incompressive condition. It is important to note that the gas remains compressible; the incompressive condition applies only to the fluctuations.

Notice from expression (\ref{DR}) that the growth rate of local modes in an equilibrium configuration is largest for modes with $k_r \ll k_\perp$. This suggests the additional approximation of restricting the analysis to modes with $\partial/\partial r \ll \ell/r$, i.e. modes that are sufficiently elongated in the radial direction that their radial variation can be neglected. I shall refer to this (along with the incompressive condition) as the \emph{oblate} limit, and the modes thus isolated as oblate modes. This approximation considerably simplifies the analysis while retaining the essential physics. In the oblate limit, which amounts to neglecting the pressure perturbation in equation (\ref{LVR}), the linear equations reduce to the set of coupled ordinary differential equations
\be\label{LO1}
\dv{\vrp}{t} = -\pdv{v_r}{r}\vrp + \frac{1}{\rhob^2}\pdv{\pb}{r}\rhop, \;\; \dv{s^\prime}{t} = -\pdv{s}{r}\vrp,
\ee
with $s^\prime = -\gamma \rhop/\rhob$. The first term on the right-hand side of the perturbed Euler equation represents angular momentum conservation: vortical modes spin faster (slower) under compression (expansion). The second term represents the baroclinic production of vorticity: entropy fluctuations spin up due to the baroclinic torque applied to them by the mean pressure gradient. The perturbed entropy equation represents the fact that, in the presence of a mean entropy gradient, entropy fluctuations evolve to compensate for mean entropy changes. An entropy fluctuation that moves up the mean entropy gradient, for example (to a region of higher mean entropy), must decrease in magnitude in order for total entropy to be conserved. This can increase or decrease the magnitude of the baroclinic term in the perturbed Euler equation.

{\color{black}One can see the connection between the perturbed Euler equation in (\ref{LO1}) and the vorticity equation as follows. The dominant vorticity component for oblate modes is the one perpendicular to both $\hat{\blds{r}}$ and $\blds{\nabla}Y_{lm}$; in axisymmetry, it is the only non-zero component and is given by
\be\label{LWPD}
\omega_\phi = \left(\blds{\nabla} \times \blds{v}^\prime \right)_\phi = \left(\frac{v_\perp^\prime - \vrp}{r} + \pdv{v_\perp^\prime}{r}\right)\pdv{P_\ell(\cos \theta)}{\theta} \approx -\frac{\vrp}{r}\pdv{P_\ell(\cos \theta)}{\theta},
\ee
where the approximation is valid for oblate modes and $P_\ell$ is a Legendre polynomial. The non-radial velocity of an oblate eddy is much smaller than its radial velocity fluctuation (this follows from $\blds{\nabla} \cdot \blds{v}^\prime \approx 0$ and $\ell/r \gg \partial/\partial r$), which implies that its vorticity is dominated by radial motion in a frame moving with the mean flow. Appendix A of \cite{Johnson14} derives the perturbed vorticity equation under the Boussinesq approximation; in spherical geometry and axisymmetry, the $\phi$ component of expression (A4) of \cite{Johnson14} (there is a sign error in front of the baroclinic term in that reference) is
\be\label{LWPE}
\dv{\omega_\phi}{t} = -\left(\frac{v_r}{r} + \pdv{v_r}{r}\right)\omega_\phi - \frac{1}{\rhob^2 r}\pdv{p}{r} \pdv{P_\ell(\cos \theta)}{\theta}\rhop
\ee
Using the approximation in (\ref{LWPD}), one can readily show that equation (\ref{LWPE}) is equivalent to the perturbed Euler equation in (\ref{LO1}).}

Using the mean flow expression $d\ln\left(\rho r^2\right)/dt = -\partial v_r/\partial r$ and recalling that the Lagrangian derivative commutes with $a$ derivatives but not $r$ derivatives, equations (\ref{LO1}) can be combined to give
\be\label{LO4}
\dv{^2 s^\prime}{t^2} + 2\pdv{v_r}{r}\dv{s^\prime}{t} + N^2 s^\prime = 0.
\ee
The only assumptions regarding the mean flow that have been made in deriving this equation are an ideal-gas equation of state and radial adiabatic flow. This equation therefore governs oblate modes in any flow satisfying these conditions. It is shown in appendix A that equation (\ref{LO4}) also applies to flows with cylindrical and planar symmetry. For $v_r = 0$, it reduces to the dispersion relation (\ref{DR}) with $k_r = 0$.

For Kidder implosions, the solutions to equation (\ref{LO4}) have the form $s^\prime \propto \exp\left(-i \int \omega \, dt\right)$, with $\omega^2 = N^2$; this demonstrates that the stability of Kidder implosions is governed by the Schwarzschild criterion. The full solution is given by
\be\label{RHOPK}
\frac{\rhop}{\rhob} = \frac{\rhop_0}{\rhob_0} \cosh \phi_K + \frac{v_{r0}^\prime}{\gamma L_{s0} \sqrt{-N_0^2}}\sinh \phi_K,
\ee
\be\label{VRPK}
\frac{\vrp}{c_s} = \frac{v_{r0}^\prime}{c_{s0}}\cosh \phi_K + \frac{\rhop_0}{\rhob_0}\frac{\gamma L_{s0} \sqrt{-N_0^2}}{c_{s0}}\sinh \phi_K,
\ee
where
\be\label{PHIK}
\phi_K \equiv \int_0^t \frac{\sqrt{-N_0^2}}{h\left(t^\prime\right)^2} \, dt^\prime = \sqrt{-N_0^2} \int_0^t \frac{dt^\prime}{1 \pm \left(t^\prime/t_c\right)^2}
\ee
and $L_{s0} \equiv r_{01}\left(\partial s_0/\partial a\right)^{-1}$ is the entropy gradient length scale at $t = 0$. In these expressions, all quantities with a zero subscript can be regarded as functions of $a$, since for oblate modes each fluid element evolves with time independently of all the others. For $N_0^2 = 0$, entropy fluctuations are conserved and $\vrp = v_{r0}^\prime/h$ (for $\rhop_0 = 0$), reflecting the conservation of angular momentum as the gas is compressed.

\begin{figure}
\hspace{0.4in}
\includegraphics[scale=0.6]{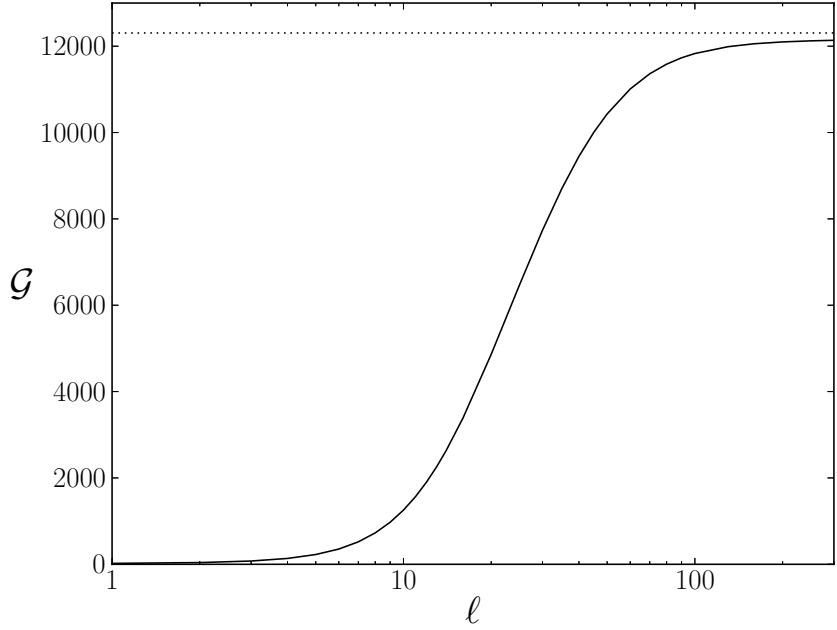}
\caption{\label{F2}Growth factor as a function of mode number for the profile shown in Fig.~\ref{F1} and a stagnating Kidder implosion/explosion (with $C = 10$), showing results from the full linear code (\emph{solid}) and the analytical oblate limit (\emph{dotted}).}
\end{figure}

\begin{figure}
\hspace{0.4in}
\includegraphics[scale=0.6]{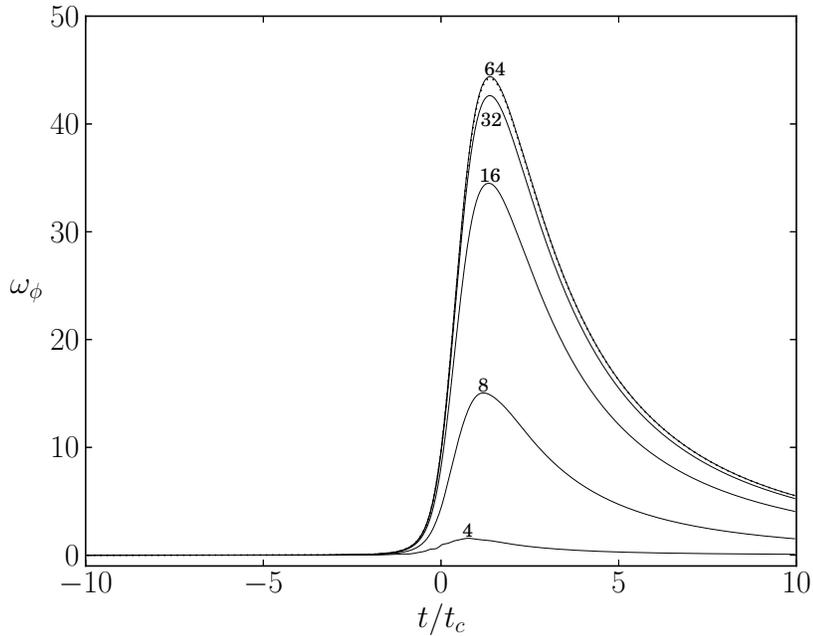}
\caption{\label{F3}{\color{black}Peak vorticity as a function of time for the profile given by expressions (\ref{RHOHS})--(\ref{THS}) (with $\sigma = 0.25$ and $a_p = 1$), a stagnating Kidder implosion/explosion (with $C = 10$), and $\ell = 30$. Shown are results from two-dimensional \texttt{Kull} calculations at several grid resolutions (\emph{solid}, labeled by zones-per-wavelength), along with results from the linear code (\emph{dotted}).}}
\end{figure}

An implosion takes place over a finite time scale, such that the impact of unstable fluctuations on the mean flow depends upon both their initial amplitude and how much they are amplified over the course of the implosion. Expression (\ref{RHOPK}) or (\ref{VRPK}) can be used to obtain an estimate of the growth factor when $N_0^2 < 0$:
\be {\cal G} \equiv \frac{\left(\rhop/\rhob\right)_{{\rm final}}}{\left(\rhop/\rhob\right)_{{\rm initial}}} \sim \left\{
\begin{array}{cc}
\left(2C\right)^{\left|N_0 t_c\right|} & \; \; {\rm (accelerating)} \\ e^{\pi\left|N_0 t_c\right|} & \; \; \, {\rm (decelerating)}. \\
\end{array}
\right.
\ee
The growth factor for the decelerating solution takes into account the explosion phase that follows the implosion phase in the Kidder stagnating solution (i.e. $h$ evolving from $C$ to $1$ and then back to $C$), since the gas is unstable both before and after stagnation. The same expressions apply to the amplification of vorticity fluctuations.

Growth factors for a representative fluid element ($a = 0.7$) undergoing a decelerating Kidder implosion/explosion are shown in Fig.~\ref{F2} as a function of Legendre mode number $\ell$. These results were obtained with a Lagrangian code that solves the full set of linear equations (\ref{LR})--(\ref{LVP}), where the radial profile of the initial entropy fluctuation was a Gaussian with a width $10\%$ of the outer radius. It can be seen from Figure~\ref{F2} that oblate modes are the fastest-growing modes and that fairly large values of $\ell$ are required to reach the oblate limit ($\ell \gtrsim 100$). {\color{black}Figure~\ref{F3} shows the evolution of the peak vorticity for $\ell = 30$ in both the linear code and a two-dimensional version of the Lagrangian hydrodynamics code \texttt{Kull} \citep{Rathkopf00} at various grid resolutions. These results indicate that a resolution of ${\sim}30$ zones per wavelength is required to accurately capture the amplification of vorticity under an implosion in a numerical simulation.}

For implosions whose trajectory is given by expression (\ref{HPL}), the solutions to equation (\ref{LO4}) again have the form $s^\prime \propto \exp\left(-i \int \omega \, dt\right)$, where here
\[
\omega^2 + \frac{5-3\gamma}{2} i D \omega - N^2 = 0
\]
and $D \equiv \partial v_r/\partial r$ is the dilatation. These solutions can equivalently be expressed as $s^\prime \propto h^{\beta}$, where
\[
\beta = -\frac{i\omega}{D} = -\frac{5-3\gamma}{4} \pm \sqrt{\Delta^2},\;\;
\Delta^2 \equiv \left(\frac{5-3\gamma}{4}\right)^2 - {\rm Ri}_c,
\]
where ${\rm Ri}_c \equiv N^2/D^2$ is a compressive Richardson number. It can be seen that instability ($\omega_{\rm i} > 0$, where $\omega_{\rm i}$ is the imaginary part of $\omega$) corresponds to $\beta_{\rm r} < 0$, where $\beta_{{\rm r}}$ is the real part of the negative branch of $\beta$ (for an implosion, a quantity that grows with time decays with $h$).

For $\gamma = 5/3$, $\beta = \pm\sqrt{-{\rm Ri}_c} = \pm\sqrt{-N_0^2 t_c^2}$, which is real for $N_0^2 < 0$. For a monatomic gas, then, the stability of these implosions is also governed by the Schwarzschild criterion. For $\gamma < 5/3$, however, the system is unconditionally unstable as a result of the additional growth due to compression. For $\Delta^2 < 0$, i.e. ${\rm Ri}_c > \left(\left[5-3\gamma\right]/4\right)^2$, the fluctuations have an oscillatory component with an amplitude that increases slowly with time. For $\Delta^2 > 0$, the growth is purely a power law in time, and increases with decreasing ${\rm Ri}_c$. The critical ${\rm Ri}_c$ above which oscillations appear varies between $0$ for $\gamma = 5/3$ and $1/4$ for $\gamma = 1$. This is reminiscent of the Richardson criterion for the stability of a stratified shear flow \citep{Miles61,Chimonas70}, with shear replaced by dilatation. In this case the transition is not between stability and instability, but rather simply between the presence and absence of oscillations.

The complete linear solution for oblate modes under implosions satisfying expression (\ref{HPL}) is
\be\label{RHOP}
\frac{\rhop}{\rhob} = h^{-\frac{5-3\gamma}{4}}\left(\frac{\rhop_0}{\rhob_0} \left[\cosh \phi + \frac{5-3\gamma}{4\Delta}\sinh \phi \right] - \frac{v_{r0}^\prime}{v_{r0}} \frac{2{\rm Ri}_c }{3[\gamma-1]\Delta}\sinh \phi \right),
\ee
\be\label{VRP}
\frac{\vrp}{v_r} = h^{-\frac{5-3\gamma}{4}} \left(\frac{v_{r0}^\prime}{v_{r0}}\left[\cosh \phi - \frac{5-3\gamma}{4\Delta} \sinh \phi\right] + \frac{\rhop_0}{\rhob_0}\frac{3[\gamma-1]}{2\Delta}\sinh \phi\right),
\ee
where $\phi \equiv \sqrt{\Delta^2} \ln h$. For $\Delta^2 < 0$, these solutions become oscillatory, and it is straightforward to show that the amplitude of the oscillations is consistent with the conservation of wave action in the fluid frame \citep{Whitham65}, i.e.
\[
\dv{}{t}\left(\frac{{\cal E}}{\omega_{{\rm r}}}\right) = 0,\;\; {\cal E} \equiv \frac{1}{2} \overline{v_r^{\prime 2}} + \frac{1}{2} N^2 \overline{\xi_r^{\prime 2}},
\]
where ${\cal E}$ is the specific energy of the fluctuations (kinetic plus potential), $\xi_r^\prime = - L_{s} s^\prime$ is the fluctuating radial fluid displacement, $\omega_{{\rm r}}$ is the real part of $\omega$, and an overbar denotes an angular average. An estimate of the growth factor for the fluctuations in this case can be made from (\ref{RHOP}) or (\ref{VRP}):
\be {\cal G} \sim C^{-\beta_{\rm r}} \sim \left\{
\begin{array}{cc}
C^{\sqrt{-{\rm Ri}_c}} \; \; \; & \;\; \;\, {\rm for} \; {\rm Ri}_c \ll -1 \\
C^{(5-3\gamma)/2} & {\rm for} \; {\rm Ri}_c = 0 \\
C^{(5-3\gamma)/4} & \; {\rm for} \; {\rm Ri}_c > 0. \\
\end{array}
\right.
\ee

\section{Summary and discussion}\label{sec:discussion}

By isolating incompressive modes that are elongated in the direction of the mean flow (oblate modes), the following conclusions are drawn regarding the stability of homologously imploding ideal gases: (i) monatomic gases are governed by the Schwarzschild criterion for buoyant stability; (ii) owing to the time-dependent nature of the mean flow, the growth is power law rather than exponential in time; (iii) additional growth occurs due to the conservation of angular momentum as vortices are compressed; and (iv) gases with $\gamma < 5/3$ are weakly unstable due to this additional growth mechanism even when the flow is otherwise buoyantly stable. As pointed out by \cite{Cao09}, the reason that \cite{Goldreich80} and \cite{Chu96} do not find instability is that they assume an isentropic background and ignore vorticity and entropy fluctuations; either of these assumptions precludes the development of buoyancy instability.

The short-wavelength nature of the most unstable modes coupled with the compression of the mean flow makes this instability challenging to capture in numerical calculations. {\color{black}Fig.~\ref{F3} indicates that accurately capturing the growth of vorticity under an implosion  requires a resolution of the order of $30$ zones per wavelength; this translates to an angular resolution of $360^\circ/(30\ell) = 12^\circ/\ell$. Capturing the fastest-growing modes ($\ell \gtrsim 100$) therefore requires an angular resolution $\lesssim 0.1^\circ$. For an Eulerian code, high convergence ratios impose severe constraints on the resolution in the direction of the mean flow.} Seeding buoyancy instability in the gas can occur in two ways: vorticity fluctuations can be transported there by shocks rippled from drive asymmetry or interfacial perturbations, or small-amplitude entropy fluctuations can be present initially in the gas. Even a sufficiently resolved calculation that is initialized without ambient density or temperature fluctuations may not capture this instability properly.

Finally, for ICF implosions and sonoluminescent bubbles, the growth at large $\ell$ is likely to be reduced by conduction and viscosity \citep{Atzeni04,Weber14}. {\color{black}These effects have been neglected here for several reasons. (i) As with classical stability analyses, a better understanding is gained if the effects driving the instability are isolated first, and stabilizing effects are added afterwards. (ii) There are uncertainties associated with conduction and viscosity models, and it is useful to know what instabilities are lurking in the background in their absence. (iii) The analytic results obtained here in the adiabatic limit provide physical insight and are useful for code verification. Self-similar solutions have been obtained previously with non-adiabatic effects \citep{Basko98,Sanz05}, although the adiabatic profiles in Fig.~\ref{F1} are remarkably similar to the actual profiles in an ICF implosion (compare Fig.~\ref{F1} with Fig.~3.11 of \citealt{Atzeni04}). Dissipation affects small scales; in reality the amplification in Fig.~\ref{F2} will fall off at high mode numbers.} At the same time, Fig.~\ref{F2} indicates that potentially significant growth can occur even for moderate $\ell$. Definitive conclusions regarding the application of this analysis to ICF and sonoluminescent bubbles will require a more faithful representation of both the mean flow and the dissipation. This, as well as application to astrophysical gases, will be pursued in future studies.
\\

I thank Dan Clark, Omar Hurricane, Karnig Mikaelian, Oleg Schilling, and the referees for their comments. This work was performed under the auspices of Lawrence Livermore National Security, LLC, (LLNS) under Contract number $\;$DE-AC52-07NA27344.

\appendix

\section{Results for cylindrical and planar symmetry}

A complete set of homologous solutions can also be derived for flows with cylindrical and planar symmetry. In general, mass and entropy conservation for homologous adiabatic flow imply $\rho(t,a) = \rho_0(a)/h^n$ and $p(t,a) = p_0(a)/h^{n\gamma}$, where $n = 1$, $2$ or $3$ for planar, cylindrical or spherical symmetry, and the time-dependent portion of (\ref{EULH}) is $h^{n\gamma-n+1}\ddot{h} = \pm t_c^{-2}$. All of the above results for the mean spatial profiles remain valid if $r$ is interpreted as the spatial coordinate in the direction of the mean flow. Expression (\ref{KIDDH}) for the Kidder implosion remains valid for $\gamma = 1+2/n$, and the generalization of (\ref{HPL}) is
\[
h(t) = \left(1 - \frac{n\gamma-n+2}{2}\sqrt{\frac{2}{n\gamma-n}} \frac{t}{t_c}\right)^\frac{2}{n\gamma-n+2},
\]
valid for $\dot{h}_0 t_c = -\sqrt{2/(n\gamma-n)}$. Both the buoyancy frequency and the dilatation have the time dependence $N \propto D \propto h^{(n-n\gamma-2)/2}$.

Equations (\ref{LO1}) for the perturbations remain valid in all three geometries, and imply
\[
\dv{^2s^\prime}{t^2} = -\pdv{s}{r}\dv{\vrp}{t} - \dv{}{t}\left(\frac{\rho r^{n-1}}{\rho_0 r_0^{n-1}}\pdv{s}{r_0}\right)\vrp,
\]
where $r_0 = r(0,a)$. Using the mean flow relation $d\ln \left(\rho r^{n-1}\right)/dt = -\partial v_r/\partial r$, this leads to equation (\ref{LO4}). The solutions for a Kidder implosion, equations (\ref{RHOPK}) and (\ref{VRPK}), are therefore valid for all three geometries (provided $\gamma = 1+2/n$). Expressions (\ref{RHOP}) and (\ref{VRP}) generalize to
\be
\frac{\rhop}{\rhob} = h^{-\frac{2+n-n\gamma}{4}}\left(\frac{\rhop_0}{\rhob_0} \left[\cosh \phi + \frac{2+n-n\gamma}{4\Delta}\sinh \phi \right] - \frac{v_{r0}^\prime}{v_{r0}} \frac{2{\rm Ri}_c }{n[\gamma-1]\Delta}\sinh \phi \right),
\ee
\be\label{VRPG}
\frac{\vrp}{v_r} = h^{-\frac{2+n-n\gamma}{4}} \left(\frac{v_{r0}^\prime}{v_{r0}}\left[\cosh \phi - \frac{2+n-n\gamma}{4\Delta} \sinh \phi\right]
+ \frac{\rhop_0}{\rhob_0}\frac{n[\gamma-1]}{2\Delta}\sinh \phi\right),
\ee
with $\Delta^2 = (2+n-n\gamma)^2/16 - {\rm Ri}_c$, so that oscillations appear for ${\rm Ri}_c > \left(2+n-n\gamma\right)^2/16$. For ${\rm Ri}_c = 0$, $\rhop_0 = 0$ and $\gamma = 5/3$, expression (\ref{VRPG}) reduces to $\vrp/v_r = \left(v_{r0}^\prime/v_{r0}\right) h^{n/3-1}$, which implies that the growth due to compression in a monatomic gas can impact the mean flow for planar and cylindrical implosions, but not for spherical implosions. In the latter case, the velocity fluctuations grow at the same rate as the mean and therefore never become large enough to drain energy from it. 

\bibliographystyle{jfm}

\bibliography{jfm-final}

\end{document}